\documentclass[12pt]{iopart}

\usepackage{iopams}

\begin{document}

\title[Density functional theory in the canonical ensemble]
{Density functional theory in the canonical ensemble: I General formalism}

\author{J A Hernando}

\address{Department of Physics, Comisi\'on Nacional de Energ\'\i a At\'omica,\\
Av. del Libertador 8250, 1429 Buenos Aires, Argentina}


\begin{abstract}
Density functional theory stems from the Hohenberg-Kohn-Sham-Mermin
(HKSM) theorem in the grand canonical ensemble (GCE). However, as recent
work shows, although its extension to the canonical ensemble (CE) is not
straightforward, work in nanopore systems could certainly benefit from a
mesoscopic DFT in the CE. The stumbling block is the fixed $N$ constraint
which is responsible for the failure in proving the interchangeability of
density profiles and external potentials as independent variables. Here we
prove that, if in the CE the correlation functions are stripped off of their
asymptotic behaviour (which is not  in the form of a properly irreducible 
$n$-body function), the HKSM theorem can be extended to the CE. In proving
that, we generate a new {\it hierarchy} of $N$-modified distribution and
correlation functions which have the same formal structure that the more
conventional ones have (but with the proper irreducible $n$-body 
behaviour) and show that, if they are employed, either a modified 
external field or the density profiles 
 can indistinctly be used as independent variables. We also
write down the $N$-modified free energy functional and prove that 
the thermodynamic potential is minimized by 
 the equilibrium values of the new hierarchy.
\end{abstract}


\pacs{61.20.Gy, 68.45.-v, 05.20.-y}

\maketitle

\section{Introduction}

Density functional theory (DFT) has become, since the seminal work of
Hohenberg, Kohn and Sham \cite{hks} and its extension to the grand canonical
ensemble (GCE) by Mermin \cite{mermin}, one of the more widely used
techniques in condensed matter theory. It has been used in a great variety
of systems \cite{revdft,evans1} as, e.g., uniform and non uniform systems in
simple \cite{kros,fmf} and general \cite{hupi} fluids, confined fluids \cite
{bryk,gevans}, melting and freezing \cite{mf}, interfaces \cite{if}, etc. as
well as in the calculation of electronic properties of all sort of systems 
\cite{revel}. This impressive amount of work has been spawned by  the
Hohenberg-Kohn-Sham-Mermin (HKSM) theorem which, roughly speaking, states
that, either the external potential or the density profile can indistinctly
be used as independent variables and that the thermodynamic grand potential
reaches its minimum value when the equilibrium density profile is used. More
precisely, it has been proved \cite{chayes} that any positive density can be
viewed as an equilibrium density for a system in a suitable external
potential. Therefore, it is in the foundation of all sort of variational
principles in the GCE. On the other hand, the failure to implement a 
straightforward canonical
ensemble (CE) extension is already well known, referred to in 
\cite{gevans} (where a GCE series 
 expansion was done and an approximate density profile
for the CE obtained), can be traced back to the fixed N constraint (see
Section 2) and it manifests itself in the failure of having an
Ornstein-Zernike (OZ) equation. In fact, years ago, Ramshaw \cite{ramshaw}
explicitly worked out the two-body OZ equation for a one component system,
showing how the stripping of the correlation function $h(r)$ off its
asymptotic behaviour solves the problems found in a more conventional
approach. Another approach to a DFT in the CE, equivalent to the one given
in \cite{gevans}, is discussed in \cite{whgonvel}. In \cite{whvel9113} the
existence of an extension of the OZ equation for the CE is discussed in the
framework of a DFT theory. As the GCE is an extension of the CE, one may
wonder about the need of an explicit CE formulation. The problem addressed
by that question is about the equivalence between both ensembles that holds
true in the thermodynamic limit but breaks down for small systems. There is
a great body of work done on the properties of fluids confined in very small
cavities which can shrink down to sizes that hold only a few adsorbed
molecules. Experiments done in porous glasses (with a mean pore
radius $r\approx 20-30$ \AA) \cite{expor} have shown how such a
degree of confinement induces strong changes with respect to bulk 
properties \cite{mf,evans90} as, e.g., critical point shifting \cite{hupi}, capillary
condensation \cite{neimark}, freezing point decreasing \cite{dominguez},
etc. Earlier approaches to these highly inhomogeneous systems included
virial expansions adapted to this situation \cite{rowlinsona,rowlinson88},
calculations of the partition function \cite{rowlinson88,gibbs} and mean
field theory \cite{gubbinsa}. Numerical simulation techniques have been also
used to study adsorption \cite{grother}, phase transitions \cite{dominguez}, 
\cite{gozdz,gubbinsb}, etc. Within the framework of DFT, the most successful
DFT that deals with inhomogeneous systems is Rosenfeld's fundamental measure
functional (FMF) \cite{fmf} which, in particular, has been used in some of
the GCE expansions referred to above (\cite{gevans},\cite{whgonvel}) in
order to obtain a CE expansion. It is quite clear, then, the interest of
having a mesoscopic DFT in the CE. Very recently we have also presented, for
one component systems, \cite{jolesdf} an extension of the HKSM theorem to
the CE.

In this paper we prove the results put forward in \cite{jolesdf} and extend
them to general mixtures. The system we consider is a $p$ components mixture
(labeled by greek indices and interacting between themselves with two-body
potentials), each $\alpha$ component
 with $N_{\alpha }$ particles in a box of volume $V$ and under
the influence of one body, species dependent, external potentials $V_{\alpha
}^{(1)}({\bf x})$. We first extend Ramshaw's work \cite{ramshaw} \ not only
by considering mixtures but, fundamentally, by proving that by stripping the
canonical correlation functions off of their asymptotic behaviour  an $N$%
-modified set of distribution and correlation function {\it hierarchies} is
generated.  Specifically, we prove that they can be defined by functional
derivatives of the canonical partition function $Q_{N}$ with respect to a
modified external potential and, moreover, that the $n$-body direct and
total correlation functions are related through $n$-body OZ\ equations. The
physical idea is the same that shows up in almost every textbook on
Statistical Mechanics when going from the CE\ to the GCE. There an open
subset is considered, here we strip the canonical correlation functions off
their asymptotic behaviour. The difference is that, for us, this is the
fundamental step, not an intermediate one. We are then able to prove that
this step is described, \ to any $n$-body order, by functional derivatives
with the right formal structure. We then prove the HKSM theorem in the CE by
introducing an $N$-modified free energy and thermodynamic potential
which is minimized by the $N$
-modified density profiles. In this way, variational principles formulated in
the CE have a firm foundation if the $N$-modified functions are used in
them. Let us emphasize that, as any successful theory of liquids needs to
consider all the correlation functions (which 
must have the correct {\it hierarchical}
structure), a valid OZ\ equation for the pair functions $h$ and $c$ is not
enough to extend DFT to the CE. Therefore, one of the crucial points of this
paper is the proof of the existence of the $N$-modified {\it hierarchies},
that they are  related by the usual formal scheme and are
linked by $n$-body OZ equations. As  side remarks, let us mention that i) the
two-body interaction potential restriction can be also lifted by including
more than one-body external potentials \cite{iyetomi} but, for the sake of
clarity, we will restrict ourselves to the two-body interaction potentials
case and ii) there exists at least one system (lattice gases) where the
conventional CE approach fails (due to the fixed topology
constraint), even in the thermodynamic limit, and it will be discussed 
in a forthcoming paper.

The layout of the paper is as follows: in Section 2 we discuss the
reasons behind the failure of a conventional CE approach, finite size
effects in the pair function $h(r)$ 
are discussed in Section 3, while in Section 4 it is proved that,
if the (already discussed) 
asymptotic behaviour of the pair canonical correlation functions is
removed, then a two-body OZ equation in the CE can be formulated. Section
5 discusses both the generation of the $N$-modified 
 distribution and correlation function
 {\it hierarchies} and of the set of $n$-body OZ
equations that relate them while, in Section 6, we introduce an $N$-modified free
energy, write down the variational equations for the density profiles and
show that the associated $N$-modified grand thermodynamic 
potential  is minimized by the equilibrium
density profiles. This completes the extension of the HKSM theorem to the CE
and, in this way, variational principles formulated in the CE have a firm
foundation if the $N$-modified functions are used in them. Section 7
summarizes our results.

\section{Failure of the canonical approach}

The system we consider is a $p$ components general mixture where the
  species are labeled by
greek indices $\alpha =1,\ldots ,p$, each one of them has $N_{\alpha }$
particles (indicated by $k_{\alpha}=1,\ldots,N_{\alpha}$) 
which are under the influence of one-body species-dependent
 external fields $V_{\alpha }^{(1)}({\bf r})$ that can be varied at
 will and the box volume is $V$. The canonical partition function is

\begin{equation}
Q(\{{\bf N}\},V,T)=\frac{1}{{\bf N}!V^{{\bf N}}}\int %
\rme ^{H_{0}+\sum_{\nu ,k_{\nu }}\Phi _{\nu }({\bf r}_{k\nu })}\rmd\{{\bf r}\}
\label{c-partf}
\end{equation}

\noindent Here $\{{\bf r}\}$denotes the configuration space of all the
particles, i.e. $\rmd\{{\bf r}\}=\prod_{\alpha k_{\alpha }}\rmd\{{\bf r}%
_{k_{\alpha }}\}$, ${\bf N}!=\prod_{\alpha =1}^{p}N_{\alpha }!$, and a $%
-\beta (=-1/kT)$ factor has been absorbed in the hamiltonian and fields. In
particular, $H_{0}$ is the system's hamiltonian in the absence of external
fields (contains all the two-body interaction terms) and $\Phi _{\alpha }(%
{\bf r})=-\beta V_{\alpha }^{(1)}({\bf r})$. The one particle canonical
distribution function is usually defined by \cite{hmcd}

\begin{eqnarray}
\fl n_{\alpha }^{(1)}({\bf x}) &= \left\langle \sum_{k_{\alpha }}\delta ({\bf x}-%
{\bf r}_{k_{\alpha }})\right\rangle =\frac{\int \sum_{k_{\alpha
}}\delta ({\bf x}-{\bf r}_{k_{\alpha }})\rme ^{H_{0}+\sum_{\nu ,k_{\nu }}\Phi
_{\nu }({\bf r}_{k\nu })}\rmd \{{\bf r}\}}{Q_{{\bf N}}}  \label{nc1-a} = \\
 \fl &=\frac{\rme ^{\Phi _{\alpha }({\bf x})}}{Q_{{\bf N}}}\frac{\delta Q_{{\bf N}}}{%
\delta \rme ^{\Phi _{\alpha }({\bf x})}}  \label{nc1-b}
\end{eqnarray}

\noindent and can also be written as

\begin{equation}
n_{\alpha }^{(1)}({\bf x})=\frac{1}{Q_{{\bf N}}}\frac{\delta Q_{{\bf N}}}{%
\delta \Phi _{\alpha }({\bf x})}=\hat{n}_{\alpha }^{(1)}({\bf x})
\label{nc1t}
\end{equation}

\noindent Both $n_{\alpha }^{(1)}({\bf x})$, the conventional canonical
distribution function, and $\hat{n}_{\alpha }^{(1)}({\bf x})$, the full
canonical distribution function (as defined by Lebowitz and Percus \cite
{leb-per1}), are the first members of two hierarchies of distribution
functions. The s-body functions of these hierarchies can be generated by
functional derivation

\begin{eqnarray}
n_{{\mbox{\boldmath $\alpha$}}}^{(s)}(\{{\bf x}\}) &=&\frac{%
\prod_{k=1}^{s}\rme ^{\Phi _{\alpha _{k}}({\bf x}_{k})}}{Q_{{\bf N}}}\frac{%
\delta ^{s}Q_{{\bf N}}}{\prod_{k=1}^{s}\delta \rme ^{\Phi _{\alpha _{k}}({\bf x}%
_{k})}}  \label{s-dens} \\
\hat{n}_{{\mbox{\boldmath $\alpha$}}}^{(s)}(\{{\bf x}\}) &=&\frac{1}{Q_{{\bf %
N}}}\frac{\delta ^{s}Q_{{\bf N}}}{\prod_{k=1}^{s}\delta \Phi _{\alpha _{k}}(%
{\bf x}_{k})}  \label{s-densf}
\end{eqnarray}

\noindent Here 
${\balpha}=(\alpha _{1},\ldots ,\alpha _{s})$%
, $\{{\bf x}\}=({\bf x}_{1},\ldots ,{\bf x}_{s})$ are the species and
coordinate sets and the difference between both hierarchies is that in the
full one different coordinates can coincide, while in the other one can
not. The GCE definition is formally the same with the grand partition
function $\Xi $ replacing $Q_{N}$. The truncated correlation functions $t$
and $\hat{t}$ associated to these hierarchies and the direct correlation
function $c$ are defined by

\begin{eqnarray}
t_{{\mbox{\boldmath $\alpha$}}}^{(s)}(\{{\bf x}\}) &=&\prod_{k=1}^{s}\rme
^{\Phi_{\alpha _{k}}({\bf x}_{k})}\frac{\delta ^{s}\ln Q_{{\bf N}}}{%
\prod_{k=1}^{s}\delta \rme ^{\Phi _{\alpha _{k}}({\bf x}_{k})}}  \label{s-trunc}
\\
\hat{t}_{{\mbox{\boldmath $\alpha$}}}^{(s)}(\{{\bf x}\}) &=&\frac{\delta
^{s}\ln Q_{{\bf N}}}{\prod_{k=1}^{s}\delta \Phi _{\alpha _{k}}({\bf x}_{k})}=%
\frac{\delta \hat{t}_{\alpha _{1}\ldots \alpha _{s-1}}^{(s-1)}({\bf x}%
_{1},\ldots ,{\bf x}_{s-1})}{\delta \Phi _{\alpha _{s}}({\bf x}_{s})}
\label{s-truncf} \\
c_{{\mbox{\boldmath $\alpha$}}}^{(s)}(\{{\bf x}\}) &=&\beta \frac{\delta
^{s}F^{(exc)}}{\prod_{k=1}^{s}\delta n_{\alpha _{k}}^{(1)}({\bf x}_{k})}=%
\frac{\delta c_{\alpha _{1}\ldots \alpha _{s-1}}^{(s-1)}({\bf x}_{1},\ldots ,%
{\bf x}_{s-1})}{\delta n_{\alpha _{s}}^{(1)}({\bf x}_{s})}  \label{s-dc}
\end{eqnarray}

\noindent $F^{(exc)}$ is the excess free energy. In particular, $t_{\alpha
}^{(1)}({\bf x})=\hat{t}_{\alpha }^{(1)}({\bf x})=n_{\alpha }^{(1)}({\bf x})=%
\hat{n}_{\alpha }^{(1)}({\bf x})$ and as

\begin{eqnarray}
\fl t_{\alpha _{1}\alpha _{2}}^{(2)}({\bf x}_{1,}{\bf x}_{2}) &=&n_{\alpha
_{1}\alpha _{2}}^{(2)}({\bf x}_{1}{\bf ,x}_{2})-n_{\alpha _{1}}^{(1)}({\bf x}%
_{1})n_{\alpha _{2}}^{(1)}({\bf x}_{2})=n_{\alpha _{1}}^{(1)}({\bf x}%
_{1})n_{\alpha _{2}}^{(1)}({\bf x}_{2})h_{\alpha _{1}\alpha _{2}}^{(2)}({\bf %
x}_{1}{\bf ,x}_{2})  \label{fh2} \\
\fl \hat{t}_{\alpha _{1}\alpha _{2}}^{(2)}({\bf x}_{1,}{\bf x}_{2}) &=&t_{\alpha
_{1}\alpha _{2}}^{(2)}({\bf x}_{1,}{\bf x}_{2})+n_{\alpha _{1}}^{(1)}({\bf x}%
_{1})\delta _{\alpha _{1}\alpha _{2}}\delta ({\bf x}_{1}{\bf -x}_{2}{\bf )}
\label{fht2}
\end{eqnarray}

\noindent the link with the more usual notation is established. Also, up to
now, the independent variables are the external fields. \noindent As by
definition is $\hat{t}_{\alpha \lambda }^{(2)}({\bf x,y)=}$ $\delta
n_{\alpha }^{(1)}({\bf x})/\delta \Phi _{\lambda }({\bf y})$ the change in
the density profiles due to changes in the external potentials is

\begin{equation}
\delta n_{\alpha }^{(1)}({\bf x})=\sum_{\lambda }\int \hat{t}_{\alpha
\lambda }^{(2)}({\bf x,y})\delta \Phi _{\lambda }({\bf y})\rmd {\bf y}
\label{dn1-dpot}
\end{equation}

\noindent If the density profiles $\{n_{\alpha }^{(1)}\}$ and the external
potentials $\{\Phi _{\lambda }\}$can interchangeably be used as independent
variables, then an inverse kernel $\hat{t}_{\eta \alpha }^{(2)-1}({\bf z,x})$
($=\delta \Phi _{\eta }({\bf z})/\delta n_{\alpha }^{(1)}({\bf x})$) must
necessarily exist such that

\begin{equation}
\sum_{\alpha }\int \hat{t}_{\eta \alpha }^{(2)-1}({\bf z,x})\hat{t}%
_{\alpha \lambda }^{(2)}({\bf x,y})\rmd {\bf x}=\delta _{\eta \lambda }\delta ({\bf z-y})
\label{ort-t}
\end{equation}

\noindent In that case

\begin{equation}
\delta \Phi _\eta ({\bf z})=\sum_\alpha \int \hat t_{\eta \alpha
}^{(2)-1}({\bf z,x})\delta n_\alpha ^{(1)}({\bf x})\rmd {\bf x}  \label{dfin1}
\end{equation}

\noindent When \ $\hat{t}_{\eta \alpha }^{(2)-1}$ is used to define the
direct correlation function $c_{\eta \alpha }^{(2)}$ \ (see equation (\ref{kerinc}%
)) then the OZ equation follows immediately.\ \ Now we can prove that, in the CE,
the inverse kernel $\hat{t}_{\eta \alpha }^{(2)-1}$ does not exist, while
 it does in the GCE.
 The normalization integrals are

\begin{equation}
\int n_{\alpha }^{(1)}({\bf x})\rmd {\bf x}=\left\{ 
\begin{tabular}{ll}
$N_{\alpha }$ & CE \\ 
$\left\langle N_{\alpha }\right\rangle $ & GCE
\end{tabular}
\right.  \label{normn1}
\end{equation}

\begin{equation}
\int n_{\alpha \lambda }^{(2)}({\bf x,y})\rmd {\bf x}\rmd {\bf y}=\left\{ 
\begin{tabular}{ll}
$N_{\alpha }(N_{\lambda }-\delta _{\alpha \lambda })$ & CE \\ 
$\left\langle N_{\alpha }(N_{\lambda }-\delta _{\alpha \lambda
})\right\rangle $ & GCE
\end{tabular}
\right.  \label{normn2}
\end{equation}
and it is easy to show that

\begin{equation}
\int n_{\lambda }^{(1)}({\bf y})\left[ 1+h_{\alpha \lambda }^{(2)}(%
{\bf x,y})\right]\rmd {\bf y} =\left\{ 
\begin{tabular}{ll}
$N_{\lambda }-\delta _{\alpha \lambda }$ & CE \\ 
$\frac{\partial \ln \left[ \Xi n_{\alpha }^{(1)}({\bf x})/z_{\alpha }\right] 
}{\partial \ln z_{\lambda }}$ & GCE
\end{tabular}
\right.  \label{normn2_b}
\end{equation}

\noindent It then follows that

\begin{equation}
\int \hat{t}_{\alpha \lambda }^{(2)}({\bf x,y})\rmd {\bf y}=\left\{ 
\begin{tabular}{ll}
$0$ & CE \\ 
$n_{\alpha }^{(1)}({\bf x})\left[ \delta _{\alpha \lambda }+\frac{\partial
\ln \left( n_{\alpha }^{(1)}({\bf x})/z_{\alpha }\right) }{\partial \ln
z_{\lambda }}\right] $ & GCE
\end{tabular}
\right.  \label{norm-t}
\end{equation}

\noindent Equations (\ref{ort-t}) and (\ref{norm-t}) are obviously incompatible
in the CE while, in the GCE and because of the fluctuations, they are not
and the kernel $\hat{t}_{\alpha \lambda }^{(2)}$ is invertible. In
consequence, the OZ equation (equations (\ref{kerinc}-\ref{ozcan})) in
terms of the conventional correlation functions does not exist in the CE. It
is also clear that the existence of an OZ equation is a previous condition
to the formulation of a DFT but, in fact, the equivalence must be established
between the existence of the whole set of $n$-body OZ\ equations and DFT
(see Sections 4 and 5).

\section{Finite size effects}

If a $\lambda $ particle is fixed in position {\bf y}, the conditional
probability of finding an $\alpha $ particle in {\bf x} is

\begin{equation}
n_{\alpha \lambda }({\bf x\mid y})=\frac{n_{\alpha \lambda }^{(2)}({\bf x,y})%
}{n_{\lambda }^{(1)}({\bf y})}=n_{\alpha }^{(1)}({\bf x})\left[ 1+h_{\alpha
\lambda }^{(2)}({\bf x,y})\right]  \label{ncond-a}
\end{equation}

\noindent This causes a density change that depends on \{$\lambda ,{\bf y}$%
\} which, when ${\bf x}$ and ${\bf y}$ are far away from each other, can
also be written as

\begin{equation}
n_{\alpha \lambda ;\infty }({\bf x\mid y})=n_{\alpha }^{(1)}({\bf x})+\frac{%
\partial n_{\alpha }^{(1)}({\bf x})}{\partial \rho _{\lambda }}\Delta \rho
_{\lambda }({\bf y)}  \label{ncond-b0}
\end{equation}

\noindent Here $\Delta \rho _{\lambda }({\bf y)}$ has the meaning of an
unknown proportionality factor which, in fact, depends on all the $\{\alpha ,%
{\bf x}\};\{\lambda ,{\bf y\}}$ variables; it is not the difference in
density that arises from fixing a $\lambda $ particle in ${\bf y}$ .
Therefore, the asymptotic behaviour of the total correlation function can be
cast in the form (uncorrelated x-y dependence)

\begin{equation}
h_{\alpha \lambda ;\infty }^{(2)}({\bf x,y)=}\frac{\partial \ln n_{\alpha
}^{(1)}({\bf x})}{\partial \rho _{\lambda }}\Delta \rho _{\lambda }({\bf y)}
\label{has-0}
\end{equation}

\noindent However, to follow this course of action (deriving with respect to
the partial density $\rho _{\lambda }$) has, at least, two inconvenients: 1) 
the $\{\alpha ,{\bf x}\}\longleftrightarrow \{\lambda ,{\bf y}\}$ exchange
symmetry would be hidden and 2) the formalism would distinguish between
diagonal ($h_{\alpha \alpha }^{(2)}$)\ and non diagonal ($h_{\alpha \lambda
}^{(2)},\alpha \neq \lambda $) correlation functions in the sense that the
long range behavior of the latter cannot be easily separated (in equation (\ref
{n1chem}) the species and field indices would be different) and, therefore,
the generalization that leads to $N$-modified hierarchy of distribution
functions is far from clear. It can also be mentioned that, while stripping
off the diagonal components is good enough for obtaining an invertible
kernel, we preferred, for the reasons mentioned above, not to do that.On the other hand, it is known \cite{lebper}
that, in the CE, the long range behaviour of the correlation 
functions is not in the nature of an
irreducible two body function, it can be written in a separated variables
form. Therefore, instead of equations  (\ref{ncond-b0}-\ref{has-0}) we 
write (redefining $\Delta \rho _{\lambda }({\bf y)}$)

\begin{equation}
h_{\alpha \lambda ;\infty }^{(2)}({\bf x,y)=}\frac{\partial \ln n_{\alpha
}^{(1)}({\bf x})}{\partial \rho _{\alpha }}\Delta \rho _{\lambda }({\bf y)}
\label{has}
\end{equation}

\noindent and similarly, for $n_{\alpha \lambda ;\infty }$. \ Defining

\begin{eqnarray*}
A_{\alpha }({\bf x)} &&{\bf =}\frac{1}{\int \frac{\partial \ln
n_{\alpha }^{(1)}({\bf u})}{\partial \rho _{\alpha }}\rmd{\bf u}}\frac{\partial \ln
n_{\alpha }^{(1)}({\bf x})}{\partial \rho _{\alpha }} \\
B_{\lambda }({\bf y)} &&{\bf =}\frac{1}{\int \Delta \rho _{\lambda }(%
{\bf v)}\rmd{\bf v}}\Delta \rho _{\lambda }({\bf y)}
\end{eqnarray*}

\noindent Both $A_{\alpha }({\bf x)}$ and $B_{\lambda }({\bf y)}$ are
normalized to one and $h_{\alpha \lambda ;\infty }^{(2)}({\bf x,y)}$ can be
written in two possible forms

\[
h_{\alpha \lambda ;\infty }^{(2)}({\bf x,y)=}\tilde{\Gamma}_{\alpha \lambda
}A_{\alpha }({\bf x)}B_{\lambda }({\bf y)=}\Gamma _{\alpha \lambda
}B_{\alpha }({\bf x)}A_{\lambda }({\bf y)} 
\]

\noindent Integrating over x we have $\tilde{\Gamma}_{\alpha \lambda
}B_{\lambda }({\bf y)=}\Gamma _{\alpha \lambda }A_{\lambda }({\bf y)}$ and so

\begin{equation}
h_{\alpha \lambda ;\infty }^{(2)}({\bf x,y)}  =\Gamma _{\alpha
\lambda }\frac{\partial \ln n_{\alpha }^{(1)}({\bf x})}{\partial \rho
_{\alpha }}\frac{\partial \ln n_{\lambda }^{(1)}({\bf y})}{\partial \rho
_{\lambda }}  \label{hasymp} \\
\end{equation}

\[
\Gamma _{\alpha \lambda } =\int \frac{\partial \ln
n_{\lambda }^{(1)}({\bf u})}{\partial \rho _{\lambda }}\rmd{\bf u}\int \Delta
\rho _{\alpha }({\bf v)\rmd{\bf v}}  \nonumber
\]

\noindent $\Gamma _{\alpha \lambda }$ depends on \{$\alpha ,\lambda $\} and
on the unknown function $\Delta \rho _{\alpha }({\bf x)}$. An $N$-modified
canonical total correlation function $\tilde{h}_{\alpha \lambda }^{(2)}$ ($N$%
-modified as implying that it is different from the correlation function $%
h_{\alpha \lambda }^{(2)}$ and that its definition is size-dependent) can
 then be defined by

\begin{equation}
h_{\alpha \lambda }^{(2)}({\bf x,y)=}\left\{ 
\begin{array}{ll}
\tilde{h}_{\alpha \lambda }^{(2)}({\bf x,y)+}\Gamma _{\alpha \lambda }\frac{%
\partial \ln n_{\alpha }^{(1)}({\bf x})}{\partial \rho _{\alpha }}\frac{%
\partial \ln n_{\lambda }^{(1)}({\bf y})}{\partial \rho _{\lambda }} & {\bf x%
}\neq {\bf y} \\ 
\tilde{h}_{\alpha \lambda }^{(2)}({\bf x,y)=-}1 & {\bf x}={\bf y}
\end{array}
\right.  \label{hstrip}
\end{equation}

\noindent The $N$-modified correlation function $\tilde{h}_{\alpha \lambda
}^{(2)}$ is the total correlation function stripped off of its asymptotic
behaviour. Notice that, although this modification seems to affect all the $%
h^{(2)}$ values, the excluded volume effects (authentically two-body) 
are not modified, not only for
the case ${\bf x}={\bf y}$ but also because, as the stripping is done
through a separation of variables, this ensures that only the long range
behaviour is affected and that the irreducible two-body component is not affected
by the stripping. Therefore, $\tilde{h}_{\alpha \lambda }^{(2)}$ is the
correlation function with a truly irreducible two-body behaviour, not $%
h_{\alpha \lambda }^{(2)}$.
  Lastly, as equation (\ref{constg-c}) shows that $\Gamma
_{\alpha \lambda }=\Gamma _{\lambda \alpha }$,\ the exchange symmetry $%
\{\alpha ,{\bf x\}\leftrightarrow }\{\lambda ,{\bf y\}}$ is clearly
preserved. The full truncated correlation function $\tilde{t}^{(2)}$
associated to $\tilde{h}^{(2)}$ is

\begin{eqnarray}
\tilde{t}_{\alpha \lambda }^{(2)}({\bf x,y}) &=&n_{\alpha }^{(1)}({\bf x}%
)n_{\lambda }^{(1)}({\bf y})\tilde{h}_{\alpha \lambda }^{(2)}({\bf x,y}%
)+n_{\alpha }^{(1)}({\bf x})\delta _{\alpha \lambda }\delta ({\bf x-y)}
\label{kercan-a} \\
&=&\left\{ 
\begin{array}{ll}
t_{\alpha \lambda }^{(2)}({\bf x,y})-\Gamma _{\alpha \lambda }\frac{\partial
n_{\alpha }^{(1)}({\bf x})}{\partial \rho _{\alpha }}\frac{\partial
n_{\lambda }^{(1)}({\bf y})}{\partial \rho _{\lambda }} & {\bf x}\neq {\bf y}
\\ 
t_{\alpha \lambda }^{(2)}({\bf x,y}) & {\bf x}={\bf y}
\end{array}
\right.  \label{kercan-b}
\end{eqnarray}

In order to write the unkwnown constant $\Gamma _{\alpha \lambda }$ as a
functional of \ $\tilde{h}_{\alpha \lambda }^{(2)}$\ let us first show that $%
\partial n_{\alpha }^{(1)}({\bf x})/\partial \rho _{\alpha }\equiv 1$. As $%
V\rho _{\alpha }=\int n_{\alpha }^{(1)}({\bf x})\rmd{\bf x}$, then $\partial
\rho _{\alpha }/\partial n_{\alpha }^{(1)}({\bf v})=1/V$ and, so, 
$\partial n_{\alpha }^{(1)}({\bf x})/\partial \rho _{\alpha }\equiv 1$.
\noindent Using \ this result in equation (\ref{norm-t}) it is obtained that

\begin{equation}
\int \tilde{t}_{\alpha \lambda }^{(2)}({\bf x,y})\rmd{\bf y}\rmd{\bf x=-}\Gamma
_{\alpha \lambda }V^{2}  \label{modker}
\end{equation}

\noindent which gets rid of the incompatibility shown in equations (\ref
{ort-t},\ref{norm-t}), allows writing the unknown constant $\Gamma
_{\alpha \lambda }$ as

\begin{equation}
\Gamma _{\alpha \lambda }=-\frac{1}{V^{2}}\int n_{\alpha }^{(1)}(%
{\bf x})\left[ \delta _{\alpha \lambda }+\int n_{\lambda }^{(1)}(%
{\bf y})\tilde{h}_{\alpha \lambda }^{(2)}({\bf x,y})\rmd {\bf y}
\right]\rmd{\bf x}  \label{constg-c}
\end{equation}

\noindent and to interpret it as a sort of average compressibility. This
recovers the classical Lebowitz-Percus results \cite{leb-per1} in a form
suitable for our purposes. We think that it is quite possible that the
factor $\Gamma _{\alpha \lambda }$ provides a formal explanation of the
changes observed in properties of confined systems with respect to bulk
properties 
\cite{expor,evans90,neimark,dominguez},
 in particular when the
system is near critical points and thus account for the observed deviation
in the critical point in confined systems \cite{hupi}. Moreover, it might be
possible that the influence of this modification becomes noticeable, even in
the thermodynamic limit, when the system is near critical points. It is also
interesting to study the normalization of the $N$-modified pair
distribution and correlation functions. Writing $h_{\alpha \lambda }^{(2)}=%
\tilde{h}_{\alpha \lambda }^{(2)}+\delta h_{\alpha \lambda }^{(2)}$ it is
obtained that (equations (\ref{hstrip}), (\ref{normn2}) and (\ref{normn2_b}))

\begin{eqnarray}
n_{\alpha }^{(1)}({\bf x})\int n_{\lambda }^{(1)}({\bf y})\delta
h_{\alpha \lambda }^{(2)}({\bf x,y})\rmd{\bf y} =\Gamma _{\alpha \lambda }V
\label{modn2-a} \\
n_{\alpha }^{(1)}({\bf x})\int n_{\lambda }^{(1)}({\bf y})\tilde{h}%
_{\alpha \lambda }^{(2)}({\bf x,y})\rmd{\bf y} =-n_{\alpha }^{(1)}({\bf x})\delta
_{\alpha \lambda }-\Gamma _{\alpha \lambda }V  \label{modn2-b} \\
\int \tilde{n}_{\alpha \lambda }^{(2)}({\bf x,y})\rmd{\bf x}\rmd{\bf y}
=N_{\alpha }(N_{\lambda }-\delta _{\alpha \lambda })-\Gamma _{\alpha
\lambda }V^{2}  \label{modn2-c}
\end{eqnarray}

\noindent and, using equation (\ref{constg-c}) for $\Gamma _{\alpha \lambda }$,
it is concluded that equation (\ref{modn2-c}) holds if

\begin{equation}
\int n_{\alpha }^{(1)}({\bf x})\rmd{\bf x}\int n_{\lambda }^{(1)}({\bf y%
})\rmd{\bf y}=N_{\alpha }N_{\lambda }  \label{n2mod-n}
\end{equation}

\noindent regardless of the value of $\int n_{\alpha }n_{\lambda }\tilde{h}%
_{\alpha \lambda }$. Therefore, the normalization of $n^{(1)}$ is enough to
ensure the normalization of $\tilde{n}^{(2)}$ or, in other words, the norm
of the $N$-modified pair correlation function is undetermined. This free
flotation effectively lifts the fixed $N$-constraint and in the next section
it is shown to be the change needed for having a two-body OZ\ equation in
the CE.

\section{Two-body OZ equation for the modified kernel}

Now we need to prove that the modified kernel $\tilde{t}_{\alpha \lambda
}^{(2)}({\bf x,y})$ can be inverted. The generalization to higher order
functions is done in Sections 5 and 6. and it is
needed not only for having a well defined $N$-modified distribution
functions hierarchy, but also for formulating a well defined CE set of OZ
equations that operate on arbitrary $n$-body configurations. If the fields
are arbitrarily varied, we get, from equations (\ref{dn1-dpot}) and (\ref
{kercan-b})

\[
\delta n_{\alpha }^{(1)}({\bf x})=\sum_{\lambda }\int \tilde{t}_{\alpha
\lambda }^{(2)}({\bf x,y})\delta \Phi _{\lambda }({\bf y})\rmd{\bf y}+%
\sum_{\lambda }\Gamma _{\alpha \lambda }\frac{\partial }{\partial \rho
_{\lambda }}\int n_{\lambda }^{(1)}({\bf y})\delta \Phi _{\lambda }({\bf y})%
\rmd{\bf y} 
\]

\noindent But equation (\ref{nc1t}) shows that, when only $\Phi _{\lambda }$
varies

\[
\delta \ln Q=-\beta \delta F=\int n_{\lambda }^{(1)}({\bf y})\delta \Phi
_{\lambda }({\bf y})\rmd{\bf y} 
\]

\noindent and, as the partial chemical potential for $\lambda $ particles is 
$\mu _{\lambda }=V^{-1}\partial F/\partial \rho _{\lambda })_{V,T}$, then

\begin{equation}
\int n_{\lambda }^{(1)}({\bf y})\delta \Phi _{\lambda }({\bf y})\rmd{\bf y}%
=-\beta V\delta \mu _{\lambda }  \label{n1chem}
\end{equation}

\noindent we can write

\[
\delta n_{\alpha }^{(1)}({\bf x})=\sum_{\lambda }\int \tilde{t}_{\alpha
\lambda }^{(2)}({\bf x,y})\delta \Phi _{\lambda }({\bf y})\rmd{\bf y}%
-\sum_{\lambda }\Gamma _{\alpha \lambda }\beta V\delta \mu _{\lambda } 
\]

\noindent and using equation (\ref{modker}) it is obtained that

\begin{eqnarray}
\delta n_{\alpha }^{(1)}({\bf x}) &=&\sum_{\lambda }\int \tilde{t}_{\alpha
\lambda }^{(2)}({\bf x,y})\delta \Phi _{\lambda }({\bf y})\rmd{\bf y}%
+\sum_{\lambda }\int \tilde{t}_{\alpha \lambda }^{(2)}({\bf x,y})\beta \delta
\mu _{\lambda }\rmd{\bf y}  \label{dn1can} \\
&=&\sum_{\lambda }\int \tilde{t}_{\alpha \lambda }^{(2)}({\bf x,y})\delta 
\tilde{\Phi}_{\lambda }({\bf y})\rmd{\bf y}
\end{eqnarray}

\noindent where $ \tilde \Phi _\lambda ({\bf y})$ is

\begin{equation}
 \tilde \Phi _\lambda ({\bf y})= \Phi _\lambda ({\bf y})+\beta
 \mu _\lambda  \label{modf}
\end{equation}

\noindent Therefore, an inverse modified kernel $\tilde{t}_{\eta \alpha
}^{(2)-1}({\bf z,x})$ exists and complies with equation (\ref{ort-t}) if the
modified field $\tilde{\Phi}$ is used instead of $\Phi $. Writing

\begin{equation}
\tilde{t}_{\eta \alpha }^{(2)-1}({\bf z,x})=\frac{\delta \tilde{\Phi}_{\eta
}({\bf z})}{\delta n_{\alpha }^{(1)}({\bf x})}=-\tilde{c}_{\eta \alpha
}^{(2)}({\bf z,x})+\frac{\delta _{\eta \alpha }\delta ({\bf z-x)}}{n_{\alpha
}^{(1)}({\bf x})}  \label{kerinc}
\end{equation}

\noindent (which defines the $N$-modified direct correlation function $%
\tilde{c}_{\eta \alpha }^{(2)}$) equation (\ref{ort-t})\ becomes the OZ\
equation in the CE where the $N$-modified pair correlation functions $\tilde{h}%
_{\eta \lambda }^{(2)}$ and $\tilde{c}_{\eta \lambda }^{(2)}$ must be used

\begin{equation}
\tilde{h}_{\eta \lambda }^{(2)}({\bf x,y})-\tilde{c}_{\eta \lambda }^{(2)}(%
{\bf x,y})=\sum_{\alpha }\int n_{\alpha }^{(1)}({\bf z})\tilde{c}%
_{\eta \alpha }^{(2)}({\bf x,z})\tilde{h}_{\alpha \lambda }^{(2)}({\bf z,y})
\rmd{\bf z}  \label{ozcan}
\end{equation}

\noindent\ So, up to now, we have seen that the constant $N$ constraint induces a long range
behavior in the conventional pair correlation function incompatible with
being an irreducible two-body function and have also shown that, if this
behaviour (expressed in the separation of variables of equation (\ref{hasymp}))
is dropped off when defining the $N$-modified pair correlation functions  (thus
getting a truly irreducible two-body function), then an authentic OZ
equation for the $N$-modified pair correlation functions is obtained. This
generalizes Ramshaw's results to pair correlations in a mixture. Let us
recall that it is essential to show that this scheme gives a coherent view
of all the possible $n$-body configurations in a liquid mixture, i.e., it is
not limited to pair functions. This is done in the next two sections.

\section{The $N$-modified canonical hierarchy}

For the sake of clarity we limit ourselves, in this section, to 
prove that we really have another 
{\it hierarchy} of distribution and correlation functions modified in such a
way that the fixed $N$\ constraint is properly taken into account and, in
the next section the new hierarchy is used to prove the 
extension of the HKSM theorem to the CE.

We first explicitly write down the modified canonical distribution function $%
\tilde{h}$ for more than two bodies and perform the same sort of analysis as
done in Sections 2 and 3 and then turn to the functional analysis
techniques in order to  show that, by deriving with respect to $\tilde{\Phi}$%
, we can generate higher order distribution functions. In this way we not
only show that functional analysis techniques are simpler but also
obtain, from the first method, 
some additional physical insight. As usual, $\tilde{h}^{(3)}$ will be
enough. First of all, the definition and normalization equations for the
three body functions are

\[
\label{cases}
\fl n_{\alpha \lambda \eta }^{(3)}({\bf x,y,z})=\cases{ 
\left\langle \sum_{k_{\alpha }}\delta ({\bf x}-{\bf r}_{k_{\alpha
}})\sum_{l_{\alpha }}\delta ({\bf y}-{\bf r}_{l_{\alpha }})\sum_{m_{\alpha
}}\delta ({\bf z}-{\bf r}_{m_{\alpha }})\right\rangle  & $\alpha \neq
\lambda \neq \eta$  \\ 
\left\langle \sum_{k_{\alpha }\neq l_{\alpha }}\delta ({\bf x}-{\bf r}%
_{k_{\alpha }})\sum_{l_{\alpha }}\delta ({\bf y}-{\bf r}_{l_{\alpha
}})\sum_{m_{\eta }}\delta ({\bf z}-{\bf r}_{m_{\eta }})\right\rangle & %
$\alpha =\lambda \neq \eta$  \\ 
\left\langle \sum_{k_{\alpha }\neq m_{\alpha }}\delta ({\bf x}-{\bf r}%
_{k_{\alpha }})\sum_{l_{\lambda }}\delta ({\bf y}-{\bf r}_{l_{\lambda
}})\sum_{m_{\alpha }}\delta ({\bf z}-{\bf r}_{m_{\alpha }})\right\rangle  & 
$\alpha =\eta \neq \lambda$  \\ 
\left\langle \sum_{k_{\alpha }}\delta ({\bf x}-{\bf r}_{k_{\alpha
}})\sum_{l_{\lambda }\neq m_{\lambda }}\delta ({\bf y}-{\bf r}_{l_{\lambda
}})\sum_{m_{\lambda }}\delta ({\bf z}-{\bf r}_{m_{\lambda }})\right\rangle 
& $\alpha \neq \lambda =\eta$  \\ 
\left\langle \sum_{k_{\alpha }\neq l_{\alpha }\neq m_{\alpha }}\delta ({\bf %
x}-{\bf r}_{k_{\alpha }})\sum_{l_{\alpha }\neq m_{\alpha }}\delta ({\bf y}-%
{\bf r}_{l_{\lambda }})\sum_{m_{\alpha }}\delta ({\bf z}-{\bf r}_{m_{\eta
}})\right\rangle  & $\alpha =\lambda =\eta$ \\}
\]

\noindent The normalization equations are, in the CE

\begin{eqnarray}
 \int n_{\alpha \lambda \eta }^{(3)}({\bf x,y,z})\rmd{\bf x}\rmd{\bf
y}\rmd{\bf z} =
N_\alpha (N_\lambda -\delta _{\alpha \lambda })(N_\eta -\delta _{\alpha \eta
}-\delta _{\lambda \eta })  \label{n3nor} \\
 \int n_\eta ^{(1)}({\bf z})h_{\alpha \lambda \eta }^{(3)}({\bf x,y,z}%
)\rmd{\bf z} =-(\delta _{\alpha \eta }+\delta _{\lambda \eta })h_{\alpha \lambda
}^{(2)}({\bf x,y})  \label{h3nor-a} \\
 \int n_\lambda ^{(1)}({\bf y})n_\eta ^{(1)}({\bf z}%
)h_{\alpha \lambda \eta }^{(3)}({\bf x,y,z})\rmd{\bf y}\rmd{\bf z} =\delta _{\alpha \lambda
}(\delta _{\alpha \eta }+\delta _{\lambda \eta })  \label{h3nor-b} \\
 \int n_\alpha ^{(1)}({\bf x})n_\lambda ^{(1)}({\bf y}%
)n_\eta ^{(1)}({\bf z})h_{\alpha \lambda \eta }^{(3)}({\bf x,y,z})
\rmd{\bf x}\rmd{\bf y}\rmd{\bf z}=N_\alpha \delta _{\alpha \lambda }(\delta _{\alpha \eta }+\delta
_{\lambda \eta })  \label{h3nor-c}
\end{eqnarray}

We can write the $N$-modified canonical correlation function $\tilde{h}%
^{(3)} $ as

\begin{equation}
h_{\alpha \lambda \eta }^{(3)}({\bf x,y,z})=\tilde{h}_{\alpha \lambda \eta
}^{(3)}({\bf x,y,z})+\delta h_{\alpha \lambda \eta }^{(3)}({\bf x,y,z})
\label{hmod-3}
\end{equation}

The interesting point to emphasize here is that, as Lebowitz and Percus \cite
{lebper} have already discussed, the long range behaviour of an $n$-body
function in the CE ensemble depends on how the clustering of the $n$
molecules is. Specifically, and in our case ($n=3$, long range 
behaviour), we can have the
three molecules very far away from each other as well as a 2-molecules
cluster and the other one very far of them. It is clear that the complexity
of the accounting for all the clustering possibilities increases
exponentially with $n$ and that, in principle, one has to define different
constants in equations like (\ref{modn2-a}) and (\ref{ct-3}), one for each
kind of clustering. This is more than enough justification to follow up the
functional differentiation approach instead of the one summarized above.
However, it is also  instructive to show that, if we consider the case of
the three molecules, every one of them very far  from each other, then it is obtained, by \
using equation (\ref{norm-t}), that

\begin{eqnarray}
\fl \int n_{\alpha }^{(1)}({\bf x})n_{\lambda }^{(1)}(%
{\bf y})n_{\eta }^{(1)}({\bf z})\delta \tilde{h}_{\alpha \lambda \eta
}^{(3)}({\bf x,y,z})\rmd{\bf x}\rmd{\bf y}\rmd{\bf z} &=&\Gamma _{\alpha \lambda \eta }V^{3}[1-\frac{\rho
_{\alpha }\delta _{\alpha \lambda }}{V}-\frac{\rho _{\lambda }\delta
_{\lambda \eta }}{V}-  \label{ct-3} \\
&&\frac{\rho _{\eta }\delta _{\alpha \eta }}{V}+\frac{\rho _{\alpha }\delta
_{\lambda \eta }}{V^{2}}+\frac{\rho _{\lambda }\delta _{\alpha \eta }}{V^{2}}%
+\frac{\rho _{\eta }\delta _{\alpha \lambda }}{V^{2}}]  \nonumber
\end{eqnarray}

\noindent If we can take the thermodynamic limit, the constant $\Gamma
_{\alpha \lambda \eta }$ is, using equation (\ref{h3nor-c}),

\begin{equation}
\fl \Gamma _{\alpha \lambda \eta }=\frac{1}{V^{3}}\int n_{\alpha }^{(1)}(%
{\bf x})\left[ \delta _{\alpha \lambda }(\delta _{\alpha \eta }+\delta
_{\lambda \eta })-\int n_{\lambda }^{(1)}({\bf y})n_{\eta
}^{(1)}({\bf z})\tilde{h}_{\alpha \lambda \eta }^{(3)}({\bf x,y,z})
\rmd{\bf y}\rmd{\bf z}\right] \rmd{\bf x}
\label{const3-b}
\end{equation}

\noindent This shows that, if the size of the system allows this kind of
limit, the long range behaviour of $n$-body configurations is (in the CE)
described by (up to)\ $n$ order susceptibilities. Lastly, let us also
mention that, working along the same lines that led to equation (\ref{n2mod-n}),
it is concluded that the norm of $h^{(3)}$ is also undetermined.

In the functional point of view, the argument is as follows: if the
densities and fields can be thought of as equivalent sets of independent
variables, i.e. given $\{n_{\alpha }\}$ the fields $\{\Phi _{\lambda }\}$
are a unique functional of them and viceversa, then each one of them can be
written as a functional Taylor series in the form

\begin{equation}
\fl \delta n_{\alpha }^{(1)}({\bf x})=\sum_{\lambda }\int \frac{\delta
n_{\alpha }^{(1)}({\bf x})}{\delta \Phi _{\lambda }({\bf y})}\delta \Phi
_{\lambda }({\bf y})\rmd{\bf y}+\frac{1}{2}\sum_{\lambda ,\eta }\int %
\frac{\delta ^{2}n_{\alpha }^{(1)}({\bf x})}{\delta \Phi _{\lambda }({\bf y}%
)\delta \Phi _{\eta }({\bf z})}\delta \Phi _{\lambda }({\bf y})\delta \Phi
_{\eta }({\bf z})\rmd{\bf y}\rmd{\bf z}+\ldots  \label{ntay-1}
\end{equation}

\begin{equation}
\fl \delta \Phi _{\alpha }({\bf x})=\sum_{\lambda }\int \frac{\delta
\Phi _{\alpha }({\bf x})}{\delta n_{\lambda }^{(1)}({\bf y})}\delta
n_{\lambda }^{(1)}({\bf y})\rmd{\bf y}+\frac{1}{2}\sum_{\lambda ,\eta }\int %
\frac{\delta ^{2}\Phi _{\alpha }({\bf x})}{\delta n_{\lambda }^{(1)}(%
{\bf y})\delta n_{\eta }^{(1)}({\bf z})}\delta n_{\lambda }^{(1)}({\bf y}%
)\delta n_{\eta }^{(1)}({\bf z})\rmd{\bf y}\rmd{\bf z}+\ldots  \label{ftay-1}
\end{equation}

\noindent Substituting $\delta \Phi $ in equation (\ref{ntay-1}) and equating
powers of $\delta n$, it is obtained, up to second order, that

\begin{equation}
\delta _{\alpha \omega }\delta ({\bf x}-{\bf u})=\sum_{\lambda }\int %
\frac{\delta n_{\alpha }^{(1)}({\bf x})}{\delta \Phi _{\lambda }({\bf y})}%
\frac{\delta \Phi _{\lambda }({\bf y})}{\delta n_{\omega }^{(1)}({\bf u})}
\rmd{\bf y}  \label{oz-1}
\end{equation}

\begin{equation}
\fl 0=\sum_{\lambda }\int \frac{\delta n_{\alpha }^{(1)}({\bf x})}{%
\delta \Phi _{\lambda }({\bf y})}\frac{\delta ^{2}\Phi _{\lambda }({\bf y})}{%
\delta n_{\omega }^{(1)}({\bf u})\delta n_{\nu }^{(1)}({\bf v})}\rmd{\bf y}%
+\sum_{\lambda ,\eta }\int \frac{\delta ^{2}n_{\alpha
}^{(1)}({\bf x})}{\delta \Phi _{\lambda }({\bf y})\delta \Phi _{\eta }({\bf z%
})}\frac{\delta \Phi _{\lambda }({\bf y})}{\delta n_{\omega }^{(1)}({\bf u})}%
\frac{\delta \Phi _{\eta }({\bf z})}{\delta n_{\nu }^{(1)}({\bf v})}
\rmd{\bf y}\rmd{\bf z}
\label{oz-2}
\end{equation}

Let us first analyze these equations in the GCE, i.e. both equations (\ref{ntay-1}%
,\ref{ftay-1}) are certainly correct. In that case, equation 
(\ref{oz-1})  just leads to the OZ
equation. Also, as equation (\ref{oz-2}) holds \noindent for all $\{\alpha ,%
{\bf x}\},\{\omega ,{\bf u}\},\{\nu ,{\bf v}\}$ values and, if we use that
(see equations (\ref{s-truncf}, \ref{kerinc}))

\begin{equation}
\hat{t}_{\alpha \lambda \eta }^{(3)}({\bf x},{\bf y,z})=\frac{\delta
^{2}n_{\alpha }^{(1)}({\bf x})}{\delta \Phi _{\lambda }({\bf y})\delta \Phi
_{\eta }({\bf z})}  \label{fh-3}
\end{equation}

\begin{eqnarray}
\hat{t}_{\alpha \lambda \eta }^{(3)-1}({\bf x},{\bf y,z}) &=&\frac{\delta
^{2}\Phi _{\alpha }({\bf x})}{\delta n_{\lambda }^{(1)}({\bf y})\delta
n_{\eta }^{(1)}({\bf z})}=\frac{\delta }{\delta n_{\lambda }^{(1)}({\bf y})}%
\left[ \hat{t}_{\alpha \eta }^{(2)-1}({\bf x,z})\right] =  \label{fhi-3} \\
&=&-\frac{\delta _{\alpha \lambda }\delta _{\alpha \eta }\delta ({\bf x}-%
{\bf y})\delta ({\bf x}-{\bf z})}{\left( n_{\alpha }^{(1)}({\bf x})\right)
^{2}}-c_{\alpha \lambda \eta }^{(3)}({\bf x},{\bf y,z})  \nonumber
\end{eqnarray}

\noindent  in equation (\ref{oz-2}), \
multiply by $\hat{t}_{\sigma \alpha }^{(2)-1}({\bf s,x})$, operate with $%
\sum_{\alpha }\int d{\bf x}$ and use equation (\ref{oz-1}), the result is

\begin{equation}
\fl \hat{t}_{\sigma \omega \nu }^{(3)-1}({\bf s,u,v})=-\sum_{\alpha ,\lambda
,\eta }\int \hat{t}_{\sigma \alpha }^{(2)-1}({\bf s,x%
})\hat{t}_{\alpha \lambda \eta }^{(3)}({\bf x},{\bf y,z})\hat{t}_{\lambda
\omega }^{(2)-1}({\bf y,u})\hat{t}_{\eta \nu }^{(2)-1}({\bf z,v})
\rmd{\bf x}\rmd{\bf y}\rmd{\bf z}    \label{oz-3}
\end{equation}

This is the OZ equation for the three-body configurations and the procedure
shows how to write OZ equations for $n$-body configurations, i.e. if the
first $p$-body OZ equations are needed, a Taylor expansion of both $%
\{n_{\alpha }^{(1)}\}$ and $\{\tilde{\Phi}_{\lambda }\}$ is performed, one
series is substituted into the other and terms up to order $p-1$ are
equated. These equations have been discussed before, see e.g. \cite
{wert,jh5,hend}, and we have followed the more straightforward Henderson's
method. Notice also that $c_{\alpha \lambda \eta }^{(3)}$ has been 
generated by density differentiation 
 of $c_{\alpha \lambda }^{(2)}$. Therefore, $\hat{t}^{(3)-1}$ exists if and only if all the $\hat{t}%
^{(2)-1}$ exist. Let us now turn to the CE. If equations (\ref{ntay-1},\ref
{ftay-1}) make sense in the CE, then in equations (\ref{oz-1},\ref{oz-2}) the
derivatives must be taken w.r.t. $\tilde{\Phi}$ and, as a consequence, $\hat{%
t}$ is replaced by $\tilde{t}$ everywhere and equation (\ref{oz-3}) (with  $%
\tilde{t}$ instead of  $\hat{t}$) is the three-body OZ equation for the $N$%
-modified correlation functions. Also, equation (\ref{fhi-3}) shows that, as
is customary, the $N$-modified direct correlation hierarchy is built by deriving
with respect to the local density. As $\hat{t}^{(n)-1}$ depends, to all
orders, on the lower order derivatives $\hat{t}^{(k)-1},k<n$, we conclude
that the operators $\delta /\delta \tilde{\Phi}$ and $\delta /\delta n$
generate the hierarchies of $N$-modified canonical total and direct
correlation functions $\tilde{t}$ and $\tilde{c}$ respectively, with $\hat{t}
$ replaced by $\tilde{t}$ and $\Phi $ by $\tilde{\Phi}$ everywhere, that the
usual formal links between both correlation functions are kept and,
lastly, let us mention that the
missing link of how the direct correlation hierarchy starts is discussed in
the next section (see equation (\ref{dirc1-c})) where we  identify the
first member of the $N$-modified direct correlation with the density
derivative of our $N$-modified free energy functional.

\section{HKSM theorem in the CE}

\bigskip Here we propose an $N$-modified free energy and thermodynamic
 potential functional, show that the latter is minimized by the equilibrium 
density profiles and
identify the first members of the $N$-modified hierarchies completing in
that way the proof that we really have a new distribution functions hierarchy with
its associated truncated and direct correlation functions hierarchies 
and that the HKSM theorem can be extended to the CE.

First of all, let us summarize the standard formulation of DFT in the GCE.
It starts from the HKSM theorem which states that, in the GCE, not only the
external potential unequivocally determines the density profile, it is also
true that the density profile unequivocally determines an external potential 
\cite{chayes}. Moreover, the grand thermodynamic potential functional is a minimum for the
equilibrium density profile. This result allows to write it 
as (following \cite{evans1} in our notation)

\begin{equation}
\beta \Omega \left[ \left\{ n^{(1)}\right\} \right] =\beta F\left[ \left\{
n^{(1)}\right\} \right] -\sum_{\alpha }\int \left[ \Phi _{\alpha }({\bf x}%
)+\beta \mu _{\alpha }\right] n_{\alpha }^{(1)}({\bf x})\rmd{\bf x}
\label{gpot-1}
\end{equation}

\noindent $F$ is the free energy functional of the system

\begin{equation}
\beta F\left[ \left\{ n^{(1)}\right\} \right] =\left\langle \beta \left(
K_{N}+U_{N}-\ln P_{N}\right) \right\rangle  \label{fenerg-1}
\end{equation}

\noindent $K_{N}$ is the kinetic energy, $U_{N}$ the interaction energy, $%
P_{N}$ the grand canonical probability density and the average is a grand
canonical one. Let us emphasize that the entropic term $\ln P_{N}$ depends
on all the $n$-body configurations and, therefore, is responsible for the
need of a full {\it hierarchy} of distribution functions. The variational
principle that determines the density profile is obtained from

\begin{equation}
\frac{\delta \beta \Omega \left[ \left\{ n^{(1)}\right\} \right] }{\delta
n_{\alpha }^{(1)}({\bf x})}=0  \label{dgpot-1}
\end{equation}

\noindent Defining the intrinsic chemical potential by

\begin{equation}
\mu _{\alpha }^{(int)}\left[ \left\{ n^{(1)}\right\} ;{\bf x}\right] =\frac{%
\delta F\left[ \left\{ n^{(1)}\right\} \right] }{\delta n_{\alpha }^{(1)}(%
{\bf x})}  \label{chemint-1}
\end{equation}

\noindent equation (\ref{dgpot-1}) becomes

\begin{equation}
\beta \mu _{\alpha }^{(int)}\left[ \left\{ n^{(1)}\right\} ;{\bf x}\right]
=\Phi _{\alpha }({\bf x})+\beta \mu _{\alpha }=\hat{\Phi}_{\alpha }({\bf x})
\label{chemint-2}
\end{equation}

\noindent where $\mu _{\alpha }$ is the chemical potential and, separating
the ideal and excess contributions to the free energy $F=F^{(id)}-F^{(exc)}$

\begin{equation}
\beta F^{(id)}=\sum_{\alpha }\int n_{\alpha }^{(1)}({\bf x})\left[ \ln
\left( \lambda ^{3}n_{\alpha }^{(1)}({\bf x})\right) -1\right] \rmd{\bf x}
\label{fenerg-id}
\end{equation}

\begin{equation}
\beta \mu _{\alpha }^{(id)}=\ln \left( \lambda ^{3}n_{\alpha }^{(1)}({\bf x}%
)\right)  \label{chemid-1}
\end{equation}

\noindent $\lambda $ is the thermal wavelength and the one body 
direct correlation
function is defined by

\begin{equation}
c_{\alpha }^{(1)}({\bf x})=\beta \frac{\delta F^{(exc)}}{\delta n_{\alpha
}^{(1)}({\bf x})}=\ln \left( \lambda ^{3}n_{\alpha }^{(1)}({\bf x})\right) -%
\hat{\Phi}_{\alpha }({\bf x}) \label{dirc1-a}
\end{equation}

\noindent which is the usual equation for the density profile \cite{evans1}.
The next direct correlation function is

\begin{equation}
c_{\alpha \lambda }^{(2)}({\bf x,y})=\frac{\delta c_{\alpha }^{(1)}({\bf x})%
}{\delta n_{\lambda }^{(1)}({\bf y})}=\frac{\delta _{\alpha \lambda }\delta (%
{\bf x-y})}{n_{\alpha }^{(1)}({\bf x})}-\frac{\delta \hat{\Phi}_{\alpha
}^{(1)}({\bf x})}{\delta n_{\lambda }^{(1)}({\bf y})}  \label{grcan-2}
\end{equation}

In this way the distribution function hierarchies are generated and a
density functional theory in the GCE is established.

Now let us turn to the CE formulation. Equations (\ref{ort-t}) and (\ref{norm-t})
show that a straightforward formulation is doomed: the inverse kernel does
not exist. In order to complete the extension of the HKSM theorem to the CE
let us write an $N$-modified free energy functional, show that there
exists a variational principle in the density profile and that a whole set
of $N$-modifed distribution function hierarchies is generated by this
functional. Our proposal is

\begin{equation}
\beta {\cal F}\left[ \left\{ \tilde{n}^{(1)}\right\} \right] =\left\langle
\beta \left( K_{N}+U_{N}-\ln P_{N}\right) \right\rangle  \label{fenerg-2}
\end{equation}

\noindent where now both $P_{N}$ and the average are canonical and the $N$%
-modified distribution functions $\tilde{n}$ are used. The $N$-modified
grand potential functional to minimize is

\begin{equation}
\beta \tilde{\Omega}\left[ \left\{ \tilde{n}^{(1)}\right\} \right] =\beta 
{\cal F}\left[ \left\{ \tilde{n}^{(1)}\right\} \right] -\sum_{\alpha }\int 
\tilde{n}_{\alpha }^{(1)}({\bf x})\tilde{\Phi}_{\alpha }({\bf x})\rmd{\bf x}
\label{fenerg-3}
\end{equation}

\noindent Let us first show that $\tilde{\Omega}$ is minimum when the
equilibrium density profiles are used in the average of equation (\ref
{fenerg-3});

\[
\fl \beta \tilde{\Omega}=\left\langle \beta \left( H-\sum_{\alpha }\mu _{\alpha
}-\ln P_{N}\right) \right\rangle =\int P_{N}\left( \{{\bf x}\}\right) \left[
\beta \left( H-\sum_{\alpha }\mu _{\alpha }-\ln P_{N}\right) \right]\rmd\{{\bf %
x}\} 
\]

\noindent and, with any other probability distribution $P_{N}^{\prime
}\left( \{{\bf x}\}\right) $ such that

\[
\int P_N^{\prime }\left( \{{\bf x}\}\right) d\{{\bf x}\}=\int P_N\left( \{%
{\bf x}\}\right) \rmd\{{\bf x}\} 
\]

\noindent the Gibbs-Bogoliubov inequality \cite{hmcd} states that

\begin{equation}
\beta \Omega ^{\prime }=\int P_{N}^{\prime }\left( \{{\bf x}\}\right) \left[
\beta \left( H-\sum_{\alpha }\mu _{\alpha }-\ln P_{N}\right) \right]\rmd\{{\bf %
x}\}\geq \beta \tilde{\Omega}  \label{fen-un}
\end{equation}

So, the variational equations are, separating as in the grand canonical case
the ideal and excess contributions,

\begin{equation}
\tilde{c}_{\alpha }^{(1)}({\bf x})=\beta \frac{\delta {\cal F}^{(exc)}}{%
\delta \tilde{n}_{\alpha }^{(1)}({\bf x})}=\ln \left( \lambda ^{3}\tilde{n}%
_{\alpha }^{(1)}({\bf x})\right) -\tilde{\Phi}_{\alpha }({\bf x})
\label{dirc1-c}
\end{equation}

\noindent This equation for the density profile has the same form as in
equation (\ref{dirc1-a}). The two body direct correlation function is

\begin{equation}
\tilde{c}_{\alpha \lambda }^{(2)}({\bf x,y})=\frac{\delta \tilde{c}_{\alpha
}^{(1)}({\bf x})}{\delta \tilde{n}_{\lambda }^{(1)}({\bf y})}=\frac{\delta
_{\alpha \lambda }\delta ({\bf x-y})}{\tilde{n}_{\alpha }^{(1)}({\bf x})}-%
\frac{\delta \tilde{\Phi}_{\alpha }^{(1)}({\bf x})}{\delta \tilde{n}%
_{\lambda }^{(1)}({\bf y})}  \label{corrc-2}
\end{equation}
and it has the same form as equation (\ref{grcan-2}). Therefore, it 
is clearly seen
that for the $N$-modified thermodynamic potential defined in equation (\ref
{fenerg-3}) a HKSM theorem can be formulated and a DFT for the CE is
established. Moreover, the thermodynamic potential of equation (\ref{fenerg-3})
shows that the $N$-modified formulation of the distribution functions allows
us to work in a CE as if it were a grand canonical one. Also, if the
system's size forces us to explicitly consider size effects, then equations like (%
\ref{kercan-b},\ref{constg-c}) contain a recipe to include size effects. A
point worth of mention is that in the first sections of this paper we wrote
the density profile as $n_{\alpha }^{(1)}$ and now, in this section, we have
used $\tilde{n}_{\alpha }^{(1)}$. In fact, both are the same ($n_{\alpha
}^{(1)}=\tilde{n}_{\alpha }^{(1)}=t_{\alpha }^{(1)}=\ldots $) but, as the CE
free energy and the $N$-modified free energy also depend on $n_{\alpha
\lambda }^{(2)}$ and $\tilde{n}_{\alpha \lambda }^{(2)}$ respectively, they
are not the same functionals and so, the hierarchies generated by functional
differentiation are different. Therefore, we use $\tilde{n}_{\alpha }^{(1)}$
just for the sake of notational consistency. 

\section{Conclusions}

Summarizing, we have proved that, if the long range behaviour of the CE
distribution and correlation functions is, roughly speaking, left aside, the
HKSM theorem and the OZ equation can be extended to the CE. As the dismissal
of the long range behaviour can be interpreted as taking an open set within
the CE\ (the textbook procedure to obtain the GCE) this seems to be an
eminently logical result. We have also generated, first by direct
construction and then by functional differentiation, a new set of $N$%
-modified distribution and correlation functions that obey the same rules
and have the same formal structure that the more conventional ones have. The
point here is that the chemical potential modifies the external field and
these new functions have the correct $n$-body irreducible behaviour. It 
can be mentioned that a price to pay for that is to give up the
normalization equations for the more than one-body correlation functions.
 In fact, only the density profile is normalized to the total number of
particles and the other functions freely float. The direct construction 
method also shows that the long range terms dropped off depend on 
$n$-body susceptibilities. We have
also shown that the functional differentiation procedure is the amenable one
when seeking a general theory \ but the direct construction method shows the
influence of the $n$-body compressibilities and is, in this sense,
physically enlightening. It might even be thought that its 
influence could be felt, even in macroscopic systems, near the
critical points but more specifric work is needed. 
 These $N$-modified functions $\tilde{n}_{\alpha
_{1}...\alpha _{s}}^{(s)}({\bf x}_{1}...{\bf x}_{s}),\tilde{t}_{\alpha
_{1}...\alpha _{s}}^{(s)}({\bf x}_{1}...{\bf x}_{s}),\tilde{c}_{\alpha
_{1}...\alpha _{s}}^{(s)}({\bf x}_{1}...{\bf x}_{s})$ are defined by

\begin{eqnarray}
\tilde{n}_{\alpha _{1}...\alpha _{s}}^{(s)}({\bf x}_{1}...{\bf x}_{s}) &=&%
\frac{1}{Q}\frac{\delta ^{s}Q}{\prod_{k=1}^{s}\delta \tilde{\Phi}_{\alpha
_{s}k}({\bf x}_{k})}  \label{modist} \\
\tilde{t}_{\alpha _{1}...\alpha _{s}}^{(s)}({\bf x}_{1}...{\bf x}_{s}) &=&%
\frac{\delta ^{s}\ln Q}{\prod_{k=1}^{s}\delta \tilde{\Phi}_{\alpha _{s}k}(%
{\bf x}_{k})}=\frac{\delta \tilde{t}_{\alpha _{1}...\alpha _{s-1}}^{(s-1)}(%
{\bf x}_{1}...{\bf x}_{s-1})}{\delta \tilde{\Phi}_{\alpha _{s}}({\bf x}_{s})}
\\
\tilde{c}_{\alpha _{1}...\alpha _{s}}^{(s)}({\bf x}_{1}...{\bf x}_{s})
&=&\beta \frac{\delta ^{s}F^{(exc)}}{\prod_{k=1}^{s}\delta \tilde{n}_{\alpha
_{k}}^{(1)}({\bf x}_{k})}=\frac{\delta \tilde{c}_{\alpha _{1}...\alpha
_{s-1}}^{(s-1)}({\bf x}_{1}...{\bf x}_{s-1})}{\delta \tilde{n}_{\alpha
_{s}}^{(1)}({\bf x}_{s})}  \label{mohcor}
\end{eqnarray}

\noindent These equations have the same structure as equations (\ref{s-dens}%
),(\ref{s-truncf}) and (\ref{s-dc}), while the direct correlation hierarchy
is initiated by 
\begin{equation}
\tilde{c}_{\alpha }^{(1)}({\bf x})=\ln \left[ \lambda ^{3}\tilde{n}_{\alpha
}^{(1)}({\bf x})\right] -\tilde{\Phi}_{\alpha }({\bf x})  \label{dircmod1}
\end{equation}

Lastly, let us also mention that, although in the introduction we have
emphasized the possible use of these results on mesoscopic systems, the next
paper in this series will deal with a system (lattice gas) 
that, interestingly enough,
needs of these results even in the thermodynamic limit.

\ack
Extremely helpful and enlightening discussions with Prof. L. Blum, which
gave rise to this work, are gratefully acknowledged.
We also acknowledge support from Fundacion Jos\'e A. Balseiro and 
from CONICET through grant PIP 0859/98.

\end{document}